\documentclass[final]{cvpr}

\usepackage{times}
\usepackage{epsfig}
\usepackage{graphicx}
\usepackage{amsmath}
\usepackage{amssymb}
\usepackage{multirow}
\usepackage{bbding}
\usepackage{pifont}
\usepackage{wasysym}
\usepackage{amssymb}

\usepackage[pagebackref=true,breaklinks=true,colorlinks,bookmarks=false]{hyperref}

\def\cvprPaperID{****} 
\def\confYear{CVPR 2021}

\begin{document}

\title{Can No-reference features help in Full-reference image quality estimation?}

\author{Saikat Dutta\\
IIT Madras\\
{\tt\small saikat.dutta779@gmail.com}
\and
Sourya Dipta Das\\
Jadavpur University\\
{\tt\small dipta.juetce@gmail.com}
\and
Nisarg A. Shah\\
IIT Jodhpur\\
{\tt\small snisarg812@gmail.com}
}

\maketitle

\begin{abstract}
Development of perceptual image quality assessment (IQA) metrics has been of significant interest to computer vision community. The aim of these metrics is to model quality of an image as perceived by humans. Recent works in 
Full-reference IQA research perform pixelwise comparison between deep features corresponding to query and reference images for quality prediction. However, pixelwise feature comparison may not be meaningful if distortion present in query image is severe. In this context, we explore utilization of no-reference features in Full-reference IQA task. Our model consists of both full-reference and no-reference branches. Full-reference branches use both distorted and reference images, whereas No-reference branch only uses distorted image. Our experiments show that use of no-reference features boosts performance of image quality assessment. Our model achieves higher SRCC and KRCC scores than a number of state-of-the-art algorithms on KADID-10K and PIPAL datasets.  
\end{abstract}

\section{Introduction}
The use of image processing to improve the quality of the content to an acceptable level for human viewers has been of substantial interest to the computer vision community. For achieving this, a significant step is to accurately measure the perceptual quality of the content, and it has been found to be critical in several computer vision applications \cite{chikkerur2011objective, sheikh2006statistical}. The primary weakness of traditional deep learning-based convolutional neural networks is that they usually generate overly smooth images instead of highly textured images because of the unavailability of the proper metrics for training these networks and quantifying these perceptual quality effects.
To an extent, Generative Adversarial Networks (GAN) \cite{goodfellow2014generative} address this problem while learning distributions with pointed peaks and adopting the adversarial training loss for image restoration tasks. Using this, they could generate sharp and visually appealing images compared to models trained without adversarial loss. Nonetheless, such adversarially trained models often obtain lower scores than those trained without adversarial loss evaluated on multiple traditional, well-known metrics such as PSNR and SSIM \cite{wang2004image}.
Efficiency of these metrics are inferior, specifically for assessing textures and fine details in the generated image \cite{ledig2017photo}. Since the ultimate goal of image enhancement models is to render visually pleasing images for humans and achieve a high Mean Opinion Score (MOS), developing a robust metric for IQA is essential.
Neural perceptual image quality metrics can also be utilized as a loss function to train deep networks for image or video restoration.


Perceptual IQA methods are of two different kinds: Full-reference methods where a distorted query image is compared against a reference image and No-reference methods where the quality of a query image is evaluated without any reference image. The common approach used in Full-reference IQA (FR-IQA) task is to extract features using a network from both reference and query images and predict perceptual quality score based on interaction ($L_2$ distance, dot product etc.) between reference and query features. However, the contribution of features extracted only from the query image in FR-IQA task is under-explored. In this paper, we propose a multi-branch model for FR-IQA problem which utilizes both Full-reference and No-reference feature extraction. In one Full-reference branch, we compute difference of ImageNet features extracted from query and reference images, whereas learnt features are compared in other Full-reference branch. No-reference branch uses only query image to extract features. Features from these three branches are concatenated and fed to fully connected layers to predict the quality score. Our model surpasses state-of-the-art FR-IQA models on two benchmark datasets.


\section{Related works}
Image Quality Assessment (IQA) methods are used to assess the quality of pictures that may have been deteriorated during processing such as generation, compression, denoising, and style transfer. IQA algorithms may be classified into no-reference and full-reference approaches based on distinct settings.
No reference methods are designed to assess image quality without the need of a reference image ~\cite{liu2017rankiqa,zhu2020metaiqa}.
On the other hand, image quality assessment in full reference setting takes in account of both distorted and reference image.

SSIM \cite{wang2004image}, MS-SSIM \cite{wang2003multiscale}, PSNR, and other full-reference approaches are extensively used. They inspired the development of FSIM \cite{zhang2011fsim}, SR-SIM \cite{zhang2012sr}, and GMSD \cite{xue2013gradient}. These hand-crafted approaches compare the feature difference between the deformed picture and the reference image to determine image quality. Deep learning-based full-reference algorithms~\cite{prashnani2018pieapp, bosse2017deep} have recently been shown to outperform hand-crafted systems in terms of model performance.

Bosse et al. \cite{bosse2017deep} implemented an architecture within a unified framework which allows joint learning of local quality and local weights. In other words, the relative importance of local quality to the final quality estimate is learnt. Their proposed architecture can be used in both No-reference (NR) and Full-reference (FR) IQA setting with minor modifications. Zhang et al.  \cite{lpips} showed deep features extracted from internal activations networks trained for high-level classification tasks, represent human perceptual similarity remarkably well, outperforming widely accepted metrics like SSIM, PSNR, FSIM and so on. They improved the performance of their network by calibrating feature outputs from a pre-trained network with their proposed BAPPS dataset. Keyan et al. \cite{Ding2020ImageQA} proposed the first full-reference IQA model with texture resampling tolerance. Their proposed method blends spatial average correlations as texture similarity with feature map correlations as structure similarity.

Prashnani et al. \cite{prashnani2018pieapp} made a large-scale dataset annotated with the probability that humans will prefer one image over another as human perceptual error. Then, using a unique pairwise-learning framework, they further proposed a new metric, PieAPP, estimated by a deep-learning model to predict which deformed image would be preferred over the other which is well correlated with human opinion. Gu et al. \cite{gu2020image} contributed Perceptual Image Processing ALgorithms (PIPAL) dataset, which is a large-scale IQA dataset. This dataset, in particular, contains the results of GAN-based Image Restoration methods, which were not included in earlier datasets. They also proposed the Space Warping Difference Network, which incorporates $L_2$ pooling layers and Space Warping Difference layers, to improve an IQA network's performance on GAN-based distortion by explicitly addressing spatial misalignment.

Cheon et al. \cite{iqt} adapt Vision transformers \cite{vit} for perceptual IQA. They have used a pretrained Inception-Resnet-v2 network \cite{inceptionresnetv2} as feature extraction backbone and transformer encoder-decoder architecture to obtain quality score predidction. Shi et al. \cite{shi2021region} proposed Region-Adaptive Deformable Network which uses reference-oriented deformable convolution to improve performance of the network on GAN-based distortion by adaptively accounting spatial misalignment. Their patch-level attention module contributes to enhance the interaction between distinct patch regions that were previously processed separately.

\begin{figure*}[!htp]
    \centering
    \includegraphics[width=0.75\textwidth]{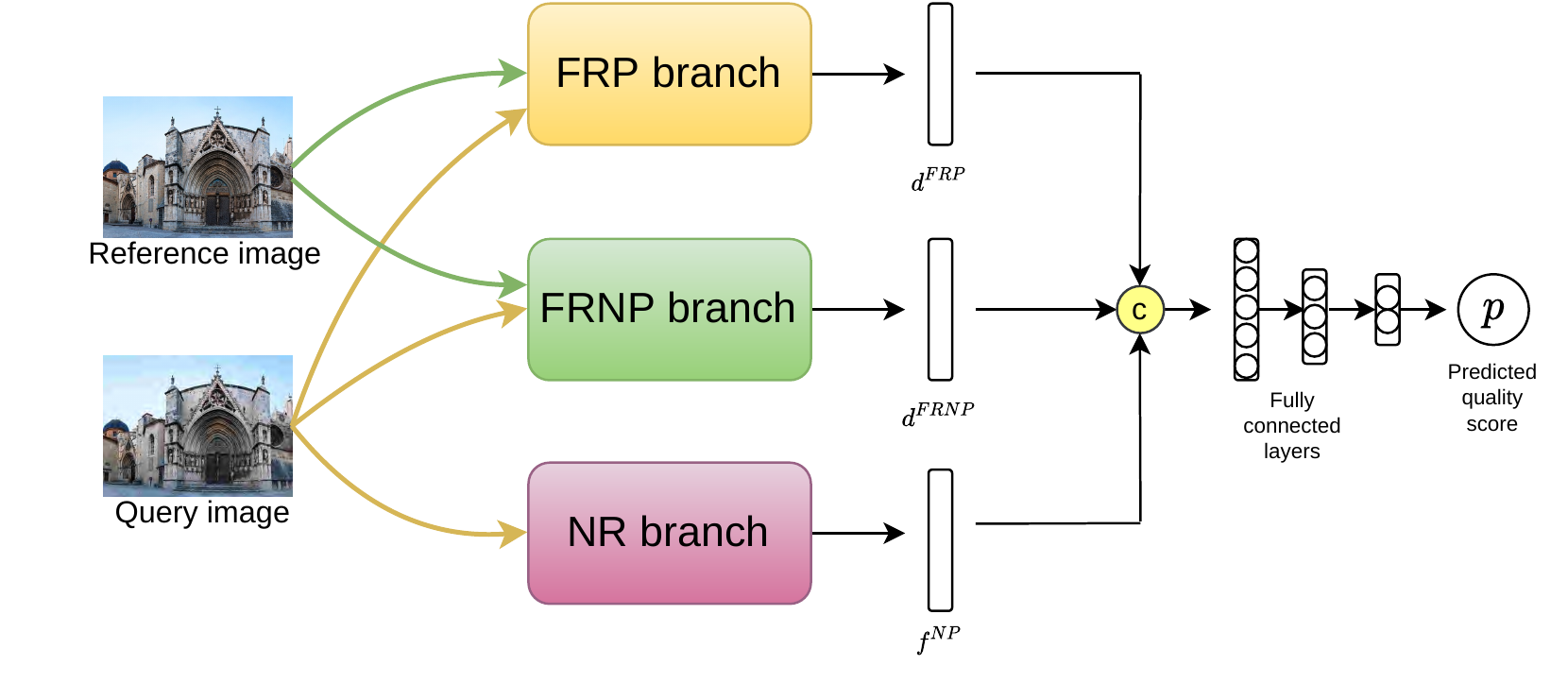}
    \caption{Overview of our FR-IQA model. Both query and reference images are fed to FRP and FRNP branches. NR branch takes only query image as input. Features from all the branches are concatenated and passed to three fully connected layers to predict the final quality score.}
    \label{fig:main}
\end{figure*}

Guo et al. \cite{guo2021iqma} sample random patches from both query and reference images and extract features from different scales from a feature extraction module. A quality score per scale is generated with the help of Feature Fusion and Score Regression modules and these scores are averaged to obtain image-level quality score.
Ayyoubzadeh et al. \cite{asna} use a Siamese-difference network equipped with spatial and channel attention. They also develop a Surrogate Ranking loss to improve Spearman Rank correlation score. 
Hammou et al. \cite{hammou2021egb} compute difference of  features extracted from different layers of pretrained VGG16 network \cite{vgg}. These features are fed to an ensemble consisting of XGBoost, LIGHTGBM and CatBoost to predict the quality score.

\section{Proposed method}

Given a reference image $I_r$ and a distorted query image $I_q$, our goal is to predict a quality score $p$ which correlates with the perceived quality of $I_q$. Our model consists of three parallel branches: (a) Full-reference pretrained (FRP) branch (b) Full-reference non-pretrained (FRNP) branch and (c) No-reference (NR) branch. 
For Full-reference branches (FRP and FRNP), both query and reference images are fed as input, whereas in No-reference branch only the query image is fed as input. 
In each of the branches, we have a convolutional neural network based encoder. In FRP branch, we use encoder from a classifier trained on ImageNet since these features are known to correlate well with perceptual quality \cite{johnson2016perceptual, dfnet, lpips}. Weights of FRP encoder is kept fixed throughout the training. We don't use pretrained encoders in FRNP and NR branches to enable learning of discriminative features from the training data. 

Full-reference branches focus on extracting features from both query and reference images and compute pixel-wise difference between query and reference features. But when the distortion is severe, computing pixel-wise difference, even in feature space, may not be optimal due to spatial misalignment. Hence, we use a No-reference branch to focus only on features related to the distortion present in the query image.

Let's denote encoders of FRP branch, FRNP branch and NR branch as $E^{FRP}$, $E^{FRNP}$ and $E^{NR}$ respectively. We describe details about Full-reference and No-reference branches in the following.


\textbf{Full-reference branches: }In a full-reference branch $b$, we extract multi-scale features from corresponding encoder $E^{b}$ for both query and reference images. We obtain features from four different scales, $\phi^{b}_{s}(I_t)$ where $ s\in\{1,2,3,4\}$ and $I_t\in\{I_q,I_r\}$. Spatial resolution of features in scale $s$ is $(h/2^s \times w/2^s)$ where $(h,w)$ is resolution of the input images. Then we compute difference features $d^b_{s}$ from each scale and concatenate them to obtain $d^b$. Hence, $d^{FRP}$ and $d^{FRNP}$ are given by,


\begin{gather}
    d^{FRP}_{s} = GAP(abs(\phi^{FRP}_{s}(I_q) - \phi^{FRP}_{s}(I_r))) \\
    d^{FRP} = d^{FRP}_1 \oplus d^{FRP}_2 \oplus d^{FRP}_3 \oplus d^{FRP}_4\\
    d^{FRNP}_{s} = GAP(abs(\phi^{FRNP}_{s}(I_q) - \phi^{FRNP}_s(I_r))) \\
    d^{FRNP} = d^{FRNP}_1 \oplus d^{FRNP}_2 \oplus d^{FRNP}_3 \oplus d^{FRNP}_4
\end{gather}

where $abs(.)$ is absolute value, $GAP(.)$ is Global Average Pooling layer and $\oplus$ stands for feature concatenation. 

\textbf{No-reference branch:} 
Since reference image is not used as an input to No-reference (NR) branch, we obtain no-reference features from this branch. Similar to Full-reference branches, multi-scale features are extracted followed by Global Average pooling and feature concatenation. 
\begin{gather}
    f^{NR}_s = GAP(\phi^{NR}_{s}(I_q))\\
    f^{NR} = f^{NR}_1(I_q) \oplus
    f^{NR}_2(I_q) \oplus
    f^{NR}_3(I_q) \oplus
    f^{NR}_4(I_q) 
\end{gather}

Finally, difference features from full-reference branches and no-reference features are concatenated and passed to three fully-connected layers to predict the quality score. Overview of our model is shown in Figure \ref{fig:main}.


\section{Experiments}

\subsection{Implementation and training details}
We implement our models with Pytorch \cite{pytorch} deep learning framework. We use Adam optimizer \cite{adam} with initial learning rate of $10^{-4}$ and batch size of 8. We gradually decrease the learning rate to $10^{-6}$. Horizontal and vertical flipping are used to augment the trainset. Mean Squared Error (MSE) loss is used as training objective. We use a machine with one NVIDIA 1080Ti GPU for our experiments. 

\subsection{Dataset Description}
We use KADID-10K \cite{lin2019kadid} and PIPAL \cite{jinjin2020pipal} datasets in our experiments.

\textbf{KADID-10K: }
KADID-10K dataset has 81 high-quality reference images of resolution $512\times384$. There are 10,125 query images of 25 traditional distortion types present in this dataset. Mean Opinion Score (MOS) range of this dataset is 1 to 5, where 1 stands for poorest quality and 5 stands for highest quality. We perform 80\%-20\% split on KADID-10K dataset for training and evaluation respectively.

\textbf{PIPAL: } PIPAL dataset consists of 250 reference images of size $288\times288$ and 29K distorted images of total 40 distortion types. This dataset contains not only traditional distortions, but also distortions produced by different image restoration algorithms including GANs. MOS scores in this dataset lies roughly within 900 and 1850, where higher score denotes better perceptual quality. We use publicly available train split of PIPAL dataset for evaluation.

\subsection{Result}
We have compared our approach against two state-of-the-art Full-reference IQA methods: WaDIQaM \cite{bosse2017deep}, LPIPS-Alex, LPIPS-VGG \cite{lpips}, and DISTS \cite{dists}. Absolute values of Spearman Rank Correlation Coefficient (SRCC) and Kendall Rank Correlation Coefficient (KRCC) are reported in Table-\ref{sota_comp} for both the datasets. Quantitative results demonstrate that our approach performs better than the other state-of-the-art methods. 

\begin{figure*}[!htp]
    \centering
    \includegraphics[width=0.8\textwidth]{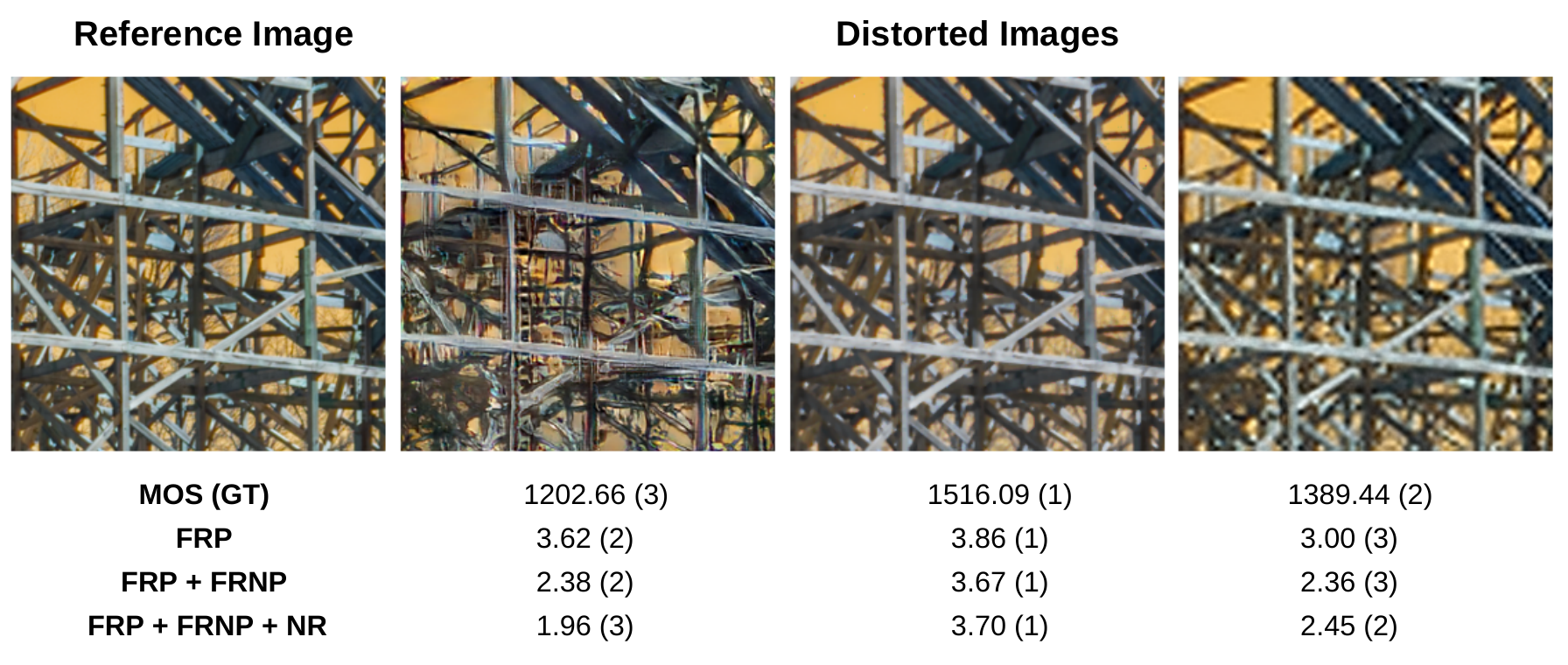}
    \caption{Effect of different branches on PIPAL dataset. Numbers within brackets denote quality-wise ranks for the corresponding method in the row. Please note, since the models are trained on KADID-10K dataset, predicted quality scores lie in range of 1-5.}
    \label{fig:abl1}
\end{figure*}

\begin{table}[!h]
\caption{Quantitative comparison with state-of-the-art methods.}
\label{sota_comp}
\small
\centering
\begin{tabular}{|c|cc|cc|}
\hline
\multirow{2}{*}{Method} & \multicolumn{2}{c|}{KADID-10K}     & \multicolumn{2}{c|}{PIPAL}         \\ \cline{2-5} 
                        & \multicolumn{1}{c|}{SRCC}   & KRCC & \multicolumn{1}{c|}{SRCC}   & KRCC \\ \hline
WaDIQaM                  & \multicolumn{1}{c|}{0.8909}       &  0.7079    & \multicolumn{1}{c|}{0.6250}       &  0.4459    \\ \hline        
LPIPS-Alex                   & \multicolumn{1}{c|}{0.8137}       &  0.6222    & \multicolumn{1}{c|}{0.5870}       &  0.4112    \\ \hline
LPIPS-VGG                   & \multicolumn{1}{c|}{0.7138}       &   0.5269   & \multicolumn{1}{c|}{0.5735}       &  0.4048    \\ \hline
DISTS                   & \multicolumn{1}{c|}{0.7966}   &   0.6094   & \multicolumn{1}{c|}{0.5785} &  0.4065    \\ \hline
Ours                    & \multicolumn{1}{c|}{\textbf{0.9536}} & \textbf{0.8128}     & \multicolumn{1}{c|}{\textbf{0.6580}} & \textbf{0.4738}     \\ \hline
\end{tabular}
\end{table}

\subsection{Ablation Study}

\textbf{Effect of different branches:} We have trained our model after removing NR and FRNP branches to understand the importance of these branches. For this experiment, we have used VGG16 backbone as feature extractor in the corresponding branches. From Table-\ref{abl_branch}, we can infer that FRP along with FRNP branch performs better than only FRP branch since FRNP branch learns dicriminative features based on different distortions present in the training data. We achieve the best performance when all three branches are used together among other configurations.
In Figure-\ref{fig:abl1}, we have shown qualitative results for different configurations on distorted images of PIPAL dataset. We can see that predictions from the full model (FRP+FRNP+NR) preserves the original quality ranking unlike other configurations.
This shows that no-reference features learnt in the NR branch aids in FR-IQA task.

\begin{table}[h]
\caption{Quantitative results for different model configurations.}
\label{abl_branch}
\centering
\footnotesize
\begin{tabular}{|c|c|c|cc|cc|}
\hline
\multirow{2}{*}{\begin{tabular}[c]{@{}c@{}}FRP\\ branch\end{tabular}} & \multirow{2}{*}{\begin{tabular}[c]{@{}c@{}}FRNP\\ branch\end{tabular}} & \multirow{2}{*}{\begin{tabular}[c]{@{}c@{}}NR\\ branch\end{tabular}} & \multicolumn{2}{c|}{KADID-10K}     & \multicolumn{2}{c|}{PIPAL}         \\ \cline{4-7} 
                                                                      &                                                                       &                                                                      & \multicolumn{1}{c|}{SRCC}   & KRCC & \multicolumn{1}{c|}{SRCC}   & KRCC \\ \hline
                                                                   \checkmark   &                                                                        &                                                                      & \multicolumn{1}{c|}{0.9200} &  0.7579    &  \multicolumn{1}{c|}{0.5650} &  0.3971    \\ \hline
                                                                   \checkmark   &        \checkmark                                                                &                                                                      & \multicolumn{1}{c|}{0.9315} &  0.7765    & \multicolumn{1}{c|}{0.6215} &  0.4422    \\ \hline
                                                                   \checkmark   &                            \checkmark                                            &                \checkmark                                                      & \multicolumn{1}{c|}{\textbf{0.9436}} & \textbf{0.7952}     & \multicolumn{1}{c|}{\textbf{0.6460}} &  \textbf{0.4617}    \\ \hline
\end{tabular}
\end{table}

\textbf{Choice of backbone:} In our experiments, we have used three different backbones in FRP, FRNP and NR branches: VGG16 \cite{vgg}, Resnet50 \cite{resnet} and Inception-v3 \cite{inception_v3}. Quantitative results are summarized in Table \ref{abl_backbone}. Our model performs the best on KADID-10K dataset when Inception-v3 is used as feature backbone, whereas we achieve best performance on PIPAL dataset when Resnet50 is used as backbone. We have chosen Inception-v3 backbone in our final model.

\begin{table}[!h]
\caption{Quantitative results for different feature extraction backbones.}
\label{abl_backbone}
\small
\centering
\begin{tabular}{|c|cc|cc|}
\hline
\multirow{2}{*}{Backbone} & \multicolumn{2}{c|}{KADID-10K}     & \multicolumn{2}{c|}{PIPAL}         \\ \cline{2-5} 
                        & \multicolumn{1}{c|}{SRCC}   & KRCC & \multicolumn{1}{c|}{SRCC}   & KRCC \\ \hline
VGG16                  & \multicolumn{1}{c|}{0.9436}       &  0.7952    & \multicolumn{1}{c|}{0.6460}       &   0.4617   \\ \hline
ResNet50                   & \multicolumn{1}{c|}{0.9493}   &   0.8068   & \multicolumn{1}{c|}{\textbf{0.6613}} &  \textbf{0.4761}    \\ \hline
Inception-v3                    & \multicolumn{1}{c|}{\textbf{0.9536}} &  \textbf{0.8128}    & \multicolumn{1}{c|}{0.6580} &  0.4738    \\ \hline
\end{tabular}
\end{table}

\section{Conclusion}
In this paper, we have proposed a multi-branch network for image quality assessment in full-reference setting. Full-reference branches computes feature from both query and reference images whereas No-reference branch extracts feature only from the query image. Full reference branches utilize both Imagenet pretrained features as well as learnt features. Our experiments show that addition of no-reference branch indeed improves FR-IQA performance. The proposed model outperforms state-of-the-art algorithms on two benchmark datasets. In future, incorporation of spatial and channel attention \cite{woo2018cbam} can be utilized to reweigh features in different branches and improve quality score predictions.

{\small
\bibliographystyle{ieee_fullname}
\bibliography{cvpr}

\begin{thebibliography}{10}\itemsep=-1pt

\bibitem{asna}
Seyed~Mehdi Ayyoubzadeh and Ali Royat.
\newblock (asna) an attention-based siamese-difference neural network with
  surrogate ranking loss function for perceptual image quality assessment.
\newblock In {\em Proceedings of the IEEE/CVF Conference on Computer Vision and
  Pattern Recognition}, pages 388--397, 2021.

\bibitem{bosse2017deep}
Sebastian Bosse, Dominique Maniry, Klaus-Robert M{\"u}ller, Thomas Wiegand, and
  Wojciech Samek.
\newblock Deep neural networks for no-reference and full-reference image
  quality assessment.
\newblock {\em IEEE Transactions on image processing}, 27(1):206--219, 2017.

\bibitem{iqt}
Manri Cheon, Sung-Jun Yoon, Byungyeon Kang, and Junwoo Lee.
\newblock Perceptual image quality assessment with transformers.
\newblock In {\em Proceedings of the IEEE/CVF Conference on Computer Vision and
  Pattern Recognition}, pages 433--442, 2021.

\bibitem{chikkerur2011objective}
Shyamprasad Chikkerur, Vijay Sundaram, Martin Reisslein, and Lina~J Karam.
\newblock Objective video quality assessment methods: A classification, review,
  and performance comparison.
\newblock {\em IEEE transactions on broadcasting}, 57(2):165--182, 2011.

\bibitem{Ding2020ImageQA}
Keyan Ding, Kede Ma, Shiqi Wang, and Eero~P. Simoncelli.
\newblock Image quality assessment: Unifying structure and texture similarity.
\newblock {\em IEEE transactions on pattern analysis and machine intelligence},
  PP, 2020.

\bibitem{dists}
K. Ding, K. Ma, S. Wang, and E.~P. Simoncelli.
\newblock Image quality assessment: Unifying structure and texture similarity.
\newblock {\em IEEE Transactions on Pattern Analysis \& Machine Intelligence},
  (01):1--1, dec 5555.

\bibitem{vit}
Alexey Dosovitskiy, Lucas Beyer, Alexander Kolesnikov, Dirk Weissenborn,
  Xiaohua Zhai, Thomas Unterthiner, Mostafa Dehghani, Matthias Minderer, Georg
  Heigold, Sylvain Gelly, et~al.
\newblock An image is worth 16x16 words: Transformers for image recognition at
  scale.
\newblock In {\em International Conference on Learning Representations}, 2020.

\bibitem{goodfellow2014generative}
Ian Goodfellow, Jean Pouget-Abadie, Mehdi Mirza, Bing Xu, David Warde-Farley,
  Sherjil Ozair, Aaron Courville, and Yoshua Bengio.
\newblock Generative adversarial nets.
\newblock {\em Advances in neural information processing systems}, 27, 2014.

\bibitem{gu2020image}
Jinjin Gu, Haoming Cai, Haoyu Chen, Xiaoxing Ye, Jimmy Ren, and Chao Dong.
\newblock Image quality assessment for perceptual image restoration: A new
  dataset, benchmark and metric.
\newblock {\em arXiv preprint arXiv:2011.15002}, 2020.

\bibitem{guo2021iqma}
Haiyang Guo, Yi Bin, Yuqing Hou, Qing Zhang, and Hengliang Luo.
\newblock Iqma network: Image quality multi-scale assessment network.
\newblock In {\em Proceedings of the IEEE/CVF Conference on Computer Vision and
  Pattern Recognition}, pages 443--452, 2021.

\bibitem{hammou2021egb}
Dounia Hammou, Sid~Ahmed Fezza, and Wassim Hamidouche.
\newblock Egb: Image quality assessment based on ensemble of gradient boosting.
\newblock In {\em Proceedings of the IEEE/CVF Conference on Computer Vision and
  Pattern Recognition}, pages 541--549, 2021.

\bibitem{resnet}
Kaiming He, Xiangyu Zhang, Shaoqing Ren, and Jian Sun.
\newblock Deep residual learning for image recognition.
\newblock In {\em Proceedings of the IEEE conference on computer vision and
  pattern recognition}, pages 770--778, 2016.

\bibitem{dfnet}
Xin Hong, Pengfei Xiong, Renhe Ji, and Haoqiang Fan.
\newblock Deep fusion network for image completion.
\newblock In {\em Proceedings of the 27th ACM international conference on
  multimedia}, pages 2033--2042, 2019.

\bibitem{jinjin2020pipal}
Gu Jinjin, Cai Haoming, Chen Haoyu, Ye Xiaoxing, Jimmy~S Ren, and Dong Chao.
\newblock Pipal: a large-scale image quality assessment dataset for perceptual
  image restoration.
\newblock In {\em European Conference on Computer Vision}, pages 633--651.
  Springer, 2020.

\bibitem{johnson2016perceptual}
Justin Johnson, Alexandre Alahi, and Li Fei-Fei.
\newblock Perceptual losses for real-time style transfer and super-resolution.
\newblock In {\em European conference on computer vision}, pages 694--711.
  Springer, 2016.

\bibitem{adam}
Diederik~P Kingma and Jimmy Ba.
\newblock Adam: A method for stochastic optimization.
\newblock {\em arXiv preprint arXiv:1412.6980}, 2014.

\bibitem{ledig2017photo}
Christian Ledig, Lucas Theis, Ferenc Husz{\'a}r, Jose Caballero, Andrew
  Cunningham, Alejandro Acosta, Andrew Aitken, Alykhan Tejani, Johannes Totz,
  Zehan Wang, et~al.
\newblock Photo-realistic single image super-resolution using a generative
  adversarial network.
\newblock In {\em Proceedings of the IEEE conference on computer vision and
  pattern recognition}, pages 4681--4690, 2017.

\bibitem{lin2019kadid}
Hanhe Lin, Vlad Hosu, and Dietmar Saupe.
\newblock Kadid-10k: A large-scale artificially distorted iqa database.
\newblock In {\em 2019 Eleventh International Conference on Quality of
  Multimedia Experience (QoMEX)}, pages 1--3. IEEE, 2019.

\bibitem{liu2017rankiqa}
Xialei Liu, Joost Van De~Weijer, and Andrew~D Bagdanov.
\newblock Rankiqa: Learning from rankings for no-reference image quality
  assessment.
\newblock In {\em Proceedings of the IEEE International Conference on Computer
  Vision}, pages 1040--1049, 2017.

\bibitem{pytorch}
Adam Paszke, Sam Gross, Francisco Massa, Adam Lerer, James Bradbury, Gregory
  Chanan, Trevor Killeen, Zeming Lin, Natalia Gimelshein, Luca Antiga, Alban
  Desmaison, Andreas Kopf, Edward Yang, Zachary DeVito, Martin Raison, Alykhan
  Tejani, Sasank Chilamkurthy, Benoit Steiner, Lu Fang, Junjie Bai, and Soumith
  Chintala.
\newblock Pytorch: An imperative style, high-performance deep learning library.
\newblock In H. Wallach, H. Larochelle, A. Beygelzimer, F. d\textquotesingle
  Alch\'{e}-Buc, E. Fox, and R. Garnett, editors, {\em Advances in Neural
  Information Processing Systems 32}, pages 8024--8035. Curran Associates,
  Inc., 2019.

\bibitem{prashnani2018pieapp}
Ekta Prashnani, Hong Cai, Yasamin Mostofi, and Pradeep Sen.
\newblock Pieapp: Perceptual image-error assessment through pairwise
  preference.
\newblock In {\em Proceedings of the IEEE Conference on Computer Vision and
  Pattern Recognition}, pages 1808--1817, 2018.

\bibitem{sheikh2006statistical}
Hamid~R Sheikh, Muhammad~F Sabir, and Alan~C Bovik.
\newblock A statistical evaluation of recent full reference image quality
  assessment algorithms.
\newblock {\em IEEE Transactions on image processing}, 15(11):3440--3451, 2006.

\bibitem{shi2021region}
Shuwei Shi, Qingyan Bai, Mingdeng Cao, Weihao Xia, Jiahao Wang, Yifan Chen, and
  Yujiu Yang.
\newblock Region-adaptive deformable network for image quality assessment.
\newblock In {\em Proceedings of the IEEE/CVF Conference on Computer Vision and
  Pattern Recognition}, pages 324--333, 2021.

\bibitem{vgg}
Karen Simonyan and Andrew Zisserman.
\newblock Very deep convolutional networks for large-scale image recognition.
\newblock {\em arXiv preprint arXiv:1409.1556}, 2014.

\bibitem{inceptionresnetv2}
Christian Szegedy, Sergey Ioffe, Vincent Vanhoucke, and Alexander~A Alemi.
\newblock Inception-v4, inception-resnet and the impact of residual connections
  on learning.
\newblock In {\em Thirty-first AAAI conference on artificial intelligence},
  2017.

\bibitem{inception_v3}
Christian Szegedy, Vincent Vanhoucke, Sergey Ioffe, Jon Shlens, and Zbigniew
  Wojna.
\newblock Rethinking the inception architecture for computer vision.
\newblock In {\em Proceedings of the IEEE conference on computer vision and
  pattern recognition}, pages 2818--2826, 2016.

\bibitem{wang2004image}
Zhou Wang, Alan~C Bovik, Hamid~R Sheikh, and Eero~P Simoncelli.
\newblock Image quality assessment: from error visibility to structural
  similarity.
\newblock {\em IEEE transactions on image processing}, 13(4):600--612, 2004.

\bibitem{wang2003multiscale}
Zhou Wang, Eero~P Simoncelli, and Alan~C Bovik.
\newblock Multiscale structural similarity for image quality assessment.
\newblock In {\em The Thrity-Seventh Asilomar Conference on Signals, Systems \&
  Computers, 2003}, volume~2, pages 1398--1402. Ieee, 2003.

\bibitem{woo2018cbam}
Sanghyun Woo, Jongchan Park, Joon-Young Lee, and In~So Kweon.
\newblock Cbam: Convolutional block attention module.
\newblock In {\em Proceedings of the European conference on computer vision
  (ECCV)}, pages 3--19, 2018.

\bibitem{xue2013gradient}
Wufeng Xue, Lei Zhang, Xuanqin Mou, and Alan~C Bovik.
\newblock Gradient magnitude similarity deviation: A highly efficient
  perceptual image quality index.
\newblock {\em IEEE transactions on image processing}, 23(2):684--695, 2013.

\bibitem{zhang2012sr}
Lin Zhang and Hongyu Li.
\newblock Sr-sim: A fast and high performance iqa index based on spectral
  residual.
\newblock In {\em 2012 19th IEEE international conference on image processing},
  pages 1473--1476. IEEE, 2012.

\bibitem{zhang2011fsim}
Lin Zhang, Lei Zhang, Xuanqin Mou, and David Zhang.
\newblock Fsim: A feature similarity index for image quality assessment.
\newblock {\em IEEE transactions on Image Processing}, 20(8):2378--2386, 2011.

\bibitem{lpips}
Richard Zhang, Phillip Isola, Alexei~A Efros, Eli Shechtman, and Oliver Wang.
\newblock The unreasonable effectiveness of deep features as a perceptual
  metric.
\newblock In {\em Proceedings of the IEEE conference on computer vision and
  pattern recognition}, pages 586--595, 2018.

\bibitem{zhu2020metaiqa}
Hancheng Zhu, Leida Li, Jinjian Wu, Weisheng Dong, and Guangming Shi.
\newblock Metaiqa: Deep meta-learning for no-reference image quality
  assessment.
\newblock In {\em Proceedings of the IEEE/CVF Conference on Computer Vision and
  Pattern Recognition}, pages 14143--14152, 2020.

\end{thebibliography}
}

\end{document}